\documentclass[11pt,twoside]{article}
\usepackage{asp2004}
\usepackage{psfig}
\usepackage{epsf}
\usepackage{graphics}
\usepackage{lscape}
\def\edcomment#1{\iffalse\marginpar{\raggedright\sl#1\/}\else\relax\fi}
\reversemarginpar

\begin{document}
\title{Properties of Type Ia Supernova Host Galaxies}
\author{P. M. Garnavich and J. Gallagher}
\affil{Physics Department, University of Notre Dame, Notre Dame, IN
46556, USA}

\begin{abstract}
We study the effect of environment on the properties of type Ia supernovae
by analyzing the integrated spectra of 57 
host galaxies.  Integrated spectra of galaxies best represent the
global properties of the host and
can be used to directly compare with spectra of high-redshift galaxies.
From the spectra we deduce the
metallicity, current star-formation rate and star-formation history of
the hosts and compare these
to the supernova decline rate which is an indicator of the luminosity.
Our results show no significant correlation between 
spiral host galaxy metallicity and SNIa light curve decline rate. The H$\alpha$
equivalent width (EW) of the host galaxy is an indicator the current star-formation rate
compared to average rate in the past. The range of SNIa luminosities increases
with decreasing EW which suggests that the variation in SNIa $^{56}$Ni production is
primarily due to progenitor mass and population age. 
\end{abstract}
\thispagestyle{plain}

\section{Introduction}

The peak luminosities and colors of type~Ia supernovae (SNIa)
appear to correlate well with their light curve decay rate \citep{P93,H96,RPK96} making
SNeIa exquisite distance indicators and powerful probes for cosmology \citep{G98,R98}.
But the origin of the diversity seen in SNIa is a mystery that limits their reliability. It is
well accepted that a SNIa occurs when a carbon-oxygen white dwarf has
accreted matter from a companion star and approaches the Chandrasekhar limit. At
some point the pressure at the center is sufficient to ignite the CO and a detonation/deflagration
unbinds the star and generates a large mass of $^{56}$Ni which powers the visible light
curve. This single-degenerate, Chandrasekhar-mass scenario predicts very consistent
explosion energies, indeed,  the peak brightness of most SNIa fall in a narrow range
spanning about 0.5 magnitudes in $B$. There is also a fraction of SNIa that are significantly
fainter than the typical event and these SNIa may represent a tail of the main distribution
or come from a completely different set of progenitors. 

One clue to the origin of diversity in SNIa is that light curve shape ({\it i.e.},peak luminosity) 
shows a dependence on host galaxy morphology as first pointed out
by \citet{H96}. Ellipticals/S0 galaxies tend to host fast-declining (low luminosity) events
compared with spiral hosts. As shown in Figure~1, the trend has become clearly established
as more SNIa have been analyzed. Galaxy morphology, however, is too blunt a tool to determine
what causes luminosity variations. For example, NGC~2841 is classified as a Sa galaxy, but
has an extremely small H$\alpha$ emission rate while NGC~632 is listed as an S0 galaxy
but has a star burst at its core. To better define the environments of the supernovae,
we have obtained integrated spectra of 57 SNIa host galaxies. From these spectra we
determine the current star-formation rate, the current rate compared to the average rate
in the past (the Scalo `b' parameter), and the metallicity for hosts with strong emission lines.

\begin{figure}[!h]
\plotfiddle{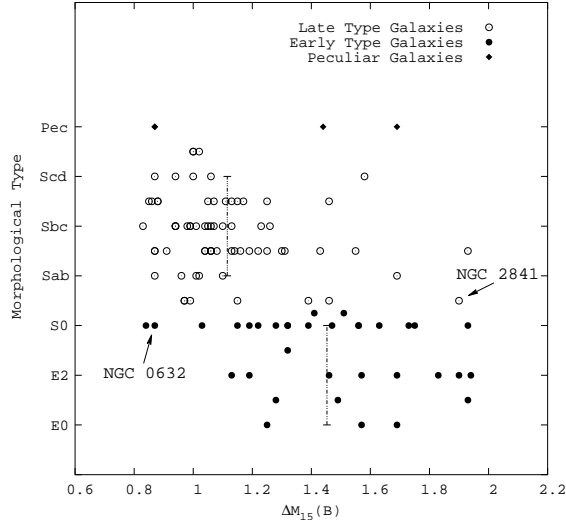}{6cm}{0}{40}{40}{-120}{-10}
\caption{The host galaxy morphology versus light curve decline rate parameter
$\Delta m_{15}(B)$.}
\end{figure}

\section{Metallicity}

The diversity in SNIa brightness comes from the range of $^{56}$Ni produced in the explosions.
The carbon to oxygen ratio in the white dwarf could effect the amount of radioactive Ni
synthesized and this ratio can be influenced by the inital metallicity of the progenitor
\citep{TBT03} or the initial mass of the progenitor \citep{Um99}. In Figure~2 we plot
the host galaxy metallicity derived from emission line ratios \citep{KD02} against
the supernova light curve decay rate. The light curve shape is parameterized by $\Delta m_{15}(B)$
which describes the number magnitudes the supernova fades 15 days after maximum light
in the $B$-band. There is no significant correlation between $\Delta m_{15}(B)$ and
host metallicity over an order of magnitude in metal content. Relying on emission lines
restricts this analysis to active star-forming galaxies, so we have supplemented our
data with elliptical galaxies published by \citet{H00}. These additional data enlarge the
range of $\Delta m_{15}(B)$ covered, but still do not reveal any correlation with metallicity.

\begin{figure}[!h]
\plotfiddle{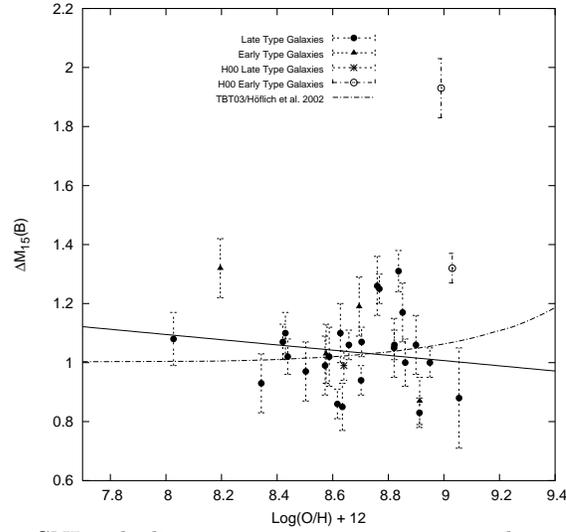}{7cm}{-90}{40}{40}{-140}{210}
\caption{SNIa decline rate parameter versus host galaxy metallicity derived from emission
line ratios. The dashed line shows the \citet{TBT03} prediction of how progenitor metallicity
effects $^{56}$Ni yield. Open symbols indicate ellipical hosts studied by \citet{H00}.}
\end{figure}

\begin{figure}[!h]
\plotfiddle{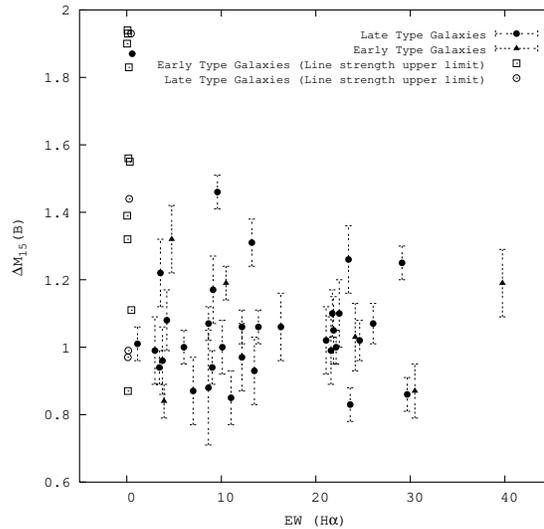}{7cm}{-90}{40}{40}{-140}{210}
\caption{SNIa decline rate parameter versus host galaxy H$\alpha$ equivalent width.}
\end{figure}

\section{Star Formation}

The rate of current star formation in a galaxy is related to its H$\alpha$ luminosity. The
ratio between current star formation and the average rate in the past can be derived from
the H$\alpha$ equivalent width. This ratio is also known as the Scalo `b' parameter and we
plot the Scalo~b versus $\Delta m_{15}(B)$ in Figure~3. For hosts with measurable emission
lines, there is no clear correlation between the decay rate and Scalo~b, but the range of SNIa
luminosity increases as the Scalo~b parameter declines. We conclude that the age of the
stellar population has more of an influence over SNIa diversity than does host metallicity.

The binary nature of the standard SNIa model means that the mass of the secondary controls
the time-scale for mass transfer \citep{Um99}. A simplified picture of the single degenerate
model is illustrated in Figure~4 where stars in the binary are selected at random from a
steep initial mass function (IMF). The primary is assumed to form a white dwarf after its main sequence
life time and explode when the secondary evolves off the main sequence if its ejected envelope is
sufficient to raise the primary white dwarf to the Chandrasekhar limit. The steep IMF means
that most SNIa come from binaries made of two low-mass stars that take 3 to 4~Gyr to explode
(as found by \citet{Strol04}). Populations older than 5~Gyr will have SNIa binaries made of a
massive primary and a low-mass secondary that takes a very long time to evolve off the main
sequence. If the $^{56}$Ni yield is inversely correlated with the primary mass, then many of the
observed properties of SNIa and their host galaxies are explained.

\begin{figure}[!h]
\plotfiddle{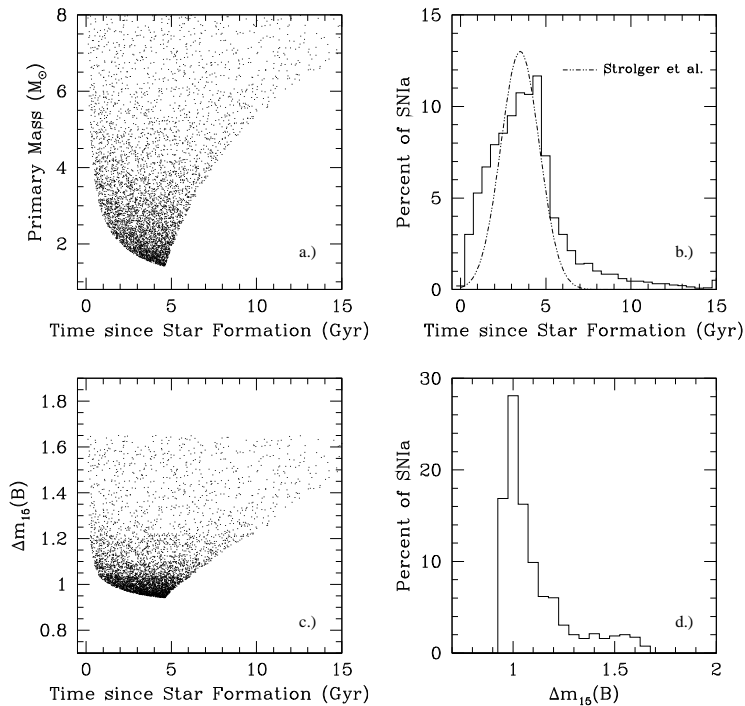}{9.0cm}{0}{50}{50}{-140}{0}
\caption{An illustration of how progenitor age (primary mass) can explain the observed
correlation between SNIa luminosity and host population. Panel $a$ shows the range of primary
masses and secondary main sequence life times that permit the primary white dwarf to reach the
Chandrasekhar limit. Panel $b$ displays the delay time for explosion after a burst of star-formation
and its similarity to the delay found in high-redshift events. Panels $c$ and $d$ show how this
model translates to SNIa light curve shape.}
\end{figure}

\end{document}